\documentclass[conference]{ieeetran}
\usepackage{amsmath}
\usepackage{graphicx}
\usepackage{cite}

\IEEEspecialpapernotice{(Invited Paper)}

\title{Autocalibrating and Calibrationless \\
Parallel Magnetic Resonance Imaging as a \\
Bilinear Inverse Problem}

\author{\IEEEauthorblockN{Martin Uecker}
\IEEEauthorblockA{
Diagnostic and Interventional Radiology\\
University Medical Center G\"ottingen,\\
G\"ottingen, Germany \\
and\\
German Center for Cardiovascular Research (DZHK)\\
partner site G\"ottingen, Germany \\
Email: martin.uecker@med.uni-goettingen.de}
}

\begin{document}

\maketitle

\begin{abstract}
Modern reconstruction methods for magnetic resonance imaging (MRI) exploit
the spatially varying sensitivity profiles of receive-coil arrays as
additional source of information. This allows to reduce the number of time-consuming
Fourier-encoding steps by undersampling. The receive sensitivities are a priori unknown 
and influenced by geometry and electric properties of the (moving) 
subject. For optimal results, they need to be estimated jointly with the image
from the same undersampled measurement data. Formulated as an inverse
problem, this leads to a bilinear  reconstruction problem related to multi-channel
blind deconvolution. In this work, we will discuss some recently
developed approaches for the solution of this problem.
\end{abstract}
\begin{IEEEkeywords}
Magnetic resonance imaging, Compressed sensing, Nonuniform sampling, Image reconstruction, Calibration
\end{IEEEkeywords}

\section{Parallel Imaging}

\begin{figure*}
\centering
\includegraphics[width=0.9\textwidth]{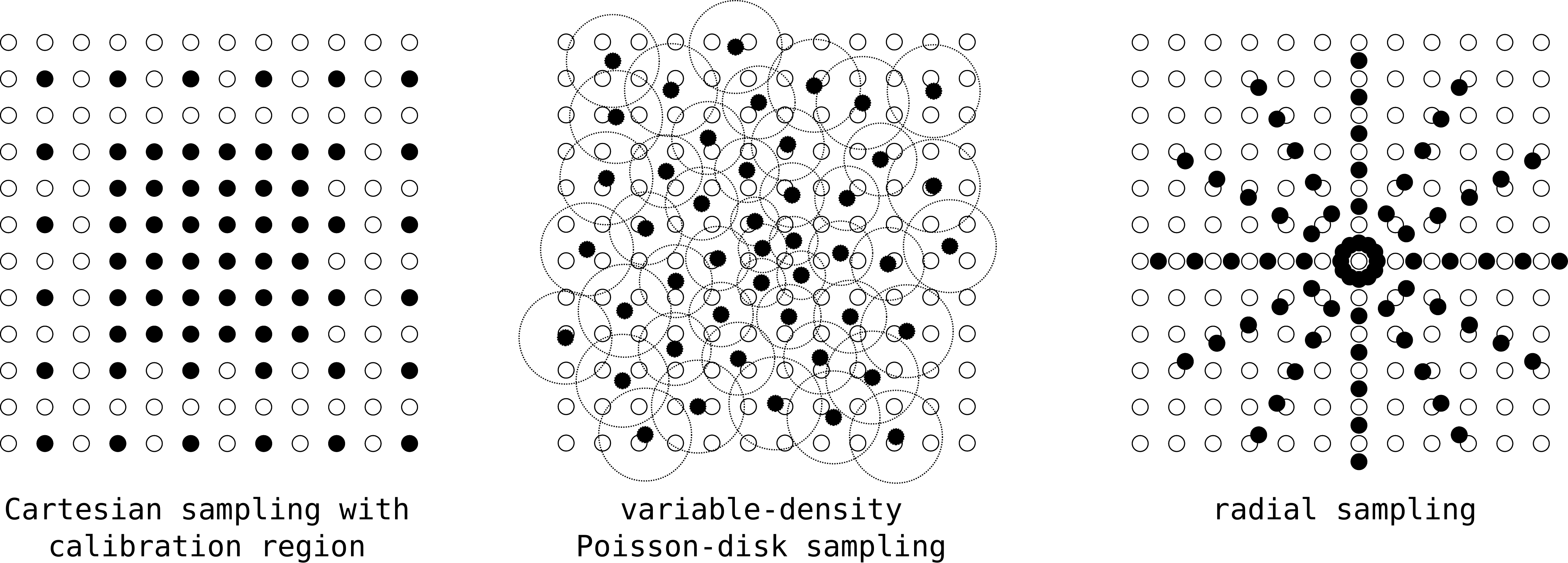}
\caption{Typical sampling schemes used in parallel MRI (from left to right):
Cartesian sampling with acceleration (undersampling) and
a fully-sampled calibration region in the k-space center, variable-density Poisson-disk sampling,
and radial sampling. Black dots represent acquired samples.} \label{fig:sampling}
\end{figure*}

\begin{figure*}
\centering
\includegraphics[width=1\textwidth]{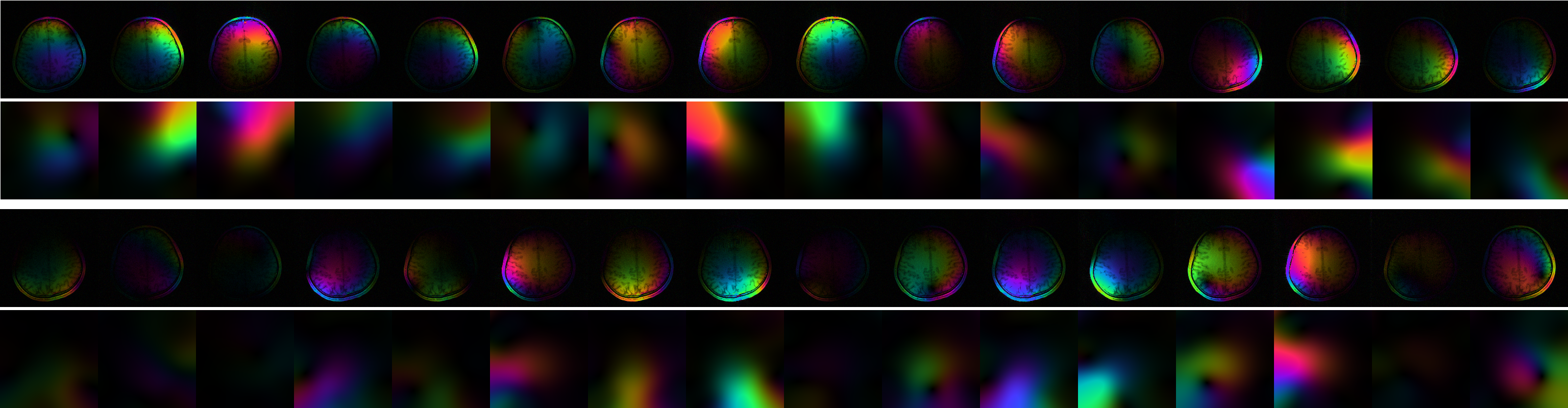}
\caption{MRI of a human brain. All 32 channels of 
a modern receiver-coil array have been reconstructed
separately from fully sampled Fourier data (top rows).
The corresponding sensitivity map of each coil 
has been estimated with the NLINV algorithm (bottom rows).
The color encodes the argument of the complex-valued maps.} 
\label{fig:channels}
\end{figure*}

The multi-channel signal $y_j$ in parallel MRI is given by the Fourier transform
of magnetization image $\rho$ weighted by the sensitivities of
the receive coils $c_j$:
\begin{align}
	y_j(\vec k) = \int_{\Omega} \textrm{d}\vec r\, \rho(\vec r) c_j(\vec r) e^{-i 2 \pi \vec k \vec r}
\end{align}
Here, $\vec k$ is the sampling trajectory in the Fourier domain (k-space),
and $\Omega$ is a compactly supported region inside $\mathcal{R}^2$ for
two-dimensional or $\mathcal{R}^3$ for three-dimensional imaging experiments. 
The extend of $\Omega$ (field-of-view) in the image domain
defines the bandwidth of the signal in k-space. If k-space
is sampled according to the corresponding Nyquist rate it is referred
to as fully sampled. As the signal equation is valid only for short durations, 
the acquisition of all required k-space samples has to be split into
a repeated series of individual measurements,
which makes the measurement process very time consuming. The main
objective in parallel MRI is to reconstruct diagnostic images while
reducing the number of samples as much as possible.
Fig.~\ref{fig:sampling} shows various common sampling schemes
and Fig.~\ref{fig:channels} shows individual coil images
and sensitivities for a modern 32-channel receiver-coil array.

The signals for all channels are highly correlated.
In fact, it can be shown that the space of ideal signals $y_j(\vec k)$
is a subspace of all band-limited k-space signals.
Parallel imaging exploits this to reduce the number
of samples below the Nyquist limit.
The subspace of ideal signals can be characterized as a Reproducing Kernel 
Hilbert Space with a matrix-valued reproducing kernel given by:
\begin{align}
	K_{st}(\vec u, \vec v) = \int_{\Omega} \textrm{d}\vec r\,  c_s(\vec r) \overline{c_t(\vec r)} e^{-i 2 \pi \left( \vec u - \vec v \right) \vec r}.
\end{align}
For known sensitivities~$c_j$, the reproducing
kernel can be computed and standard approximation theory then
yields a complete sampling theory with optimal
formulas for interpolation and error bounds~\cite{RKHS}.

More commonly, image reconstruction is
formulated as a linear inverse problem and
a regularized solution is defined as the minimizer of
the following optimization problem \cite{CGSENSE,UeckerCRC15}:
\begin{align}
	\underset{x}{\operatorname{arg min}} \sum_j \|P \mathcal{F} S_j x - y_j \|_2^2 + R(x)
\end{align}
Here, $S_j$ is the multiplication with the sensitivity $c_j$, $\mathcal{F}$
the Fourier transform, and $P$ the sampling operator, and $R(x)$ a
regularization term.
Most state-of-the art algorithms for MRI are based on numerical
optimization of the discretized functional and often also combine
parallel imaging with concepts from compressed
sensing to further reduce the number of samples \cite{SparseMRI,TVMRI}.

\section{Autocalibrating Parallel Imaging}

In practice, the coil sensitivities~$c_j$ are not known.
Many techniques have been developed for autocalibration
of the sensitivities from a fully-sampled 
(calbration) region $y_{calib}$ in the center of k-space
parallel imaging. Early schemes estimated
the coil sensitivities directly from low-resolution images
obtained as the inverse Fourier transform of the
calibration region, but
these methods are not optimal and need a high number
of additional samples. In the following,
we will discuss ESPIRiT as a recent subspace-based approach
for autocalibration~\cite{ESPIRiT}.

Calibration can be formulated using the
calibration matrix $\mathcal{C}\left\{y_{calib}\right\}$, which
is constructed as a block-Hankel matrix from fully sampled
calibration data $y_{calib}$ by taking
overlapping patches of the multi-channel k-space as
rows. Because patches are locally correlated in
k-space between receive channels, this matrix has a 
low rank and singular-value decomposition can be
used to compute the basis for the signal subspace 
$V_{\parallel}$ and for the noise subspace $V_{\perp}$
of patches in k-space.

Given a calibration matrix, highly accurate
sensitivities can be recovered using the ESPIRiT algorithm:
The null-space condition $V_{\perp}V_{\perp}^H R_r \hat y = 0$
for patches $R_r \hat y$ of the unknown k-space $\hat y$
($r$ enumerating all patches) yields an overdetermined system
of  equations which can be solved in the least-squares sense
using the normal equations:
\begin{align}
	\sum_r R_r^H V_{\perp} V_{\perp}^H R_r \hat y & = 0 
\end{align}
This can be rewritten as a
a matrix-valued convolution $\mathcal{W}$ that
reproduces an ideal fully-sampled k-space ($M$ is the
number of samples in a patch):
\begin{align}
	\underbrace{M^{-1} \sum_r R_r^H  V_{\parallel} V_{\parallel}^H R_r}_{\mathcal{W}} y & = y 
\end{align}
Combining this equation $\mathcal{W} y = y$ with the ideal
signal model $y = \mathcal{F} S \rho$ and applying an
inverse Fourier transform yields $\mathcal{F}^{-1} \mathcal{W} \mathcal{F} S = 1 S$ where $\rho \neq 0$.
In other words, the coil sensitivities appear as eigenvectors to the eigenvalue one
of the operator $\mathcal{F}^{-1} \mathcal{W} \mathcal{F}$. As this
operator is point-wise in the image domain, it can be efficiently
be computed.
Interestingly, in the case of certain inconsistencies in the data
the algorithm produces multiple sets of eigenvector maps to the
eigenvalue one. This can be exploited for more robust parallel
MRI by using an extended optimization problem for image reconstruction:
\begin{align}
	\underset{x^i}{\operatorname{arg min}} \sum_j \|P \mathcal{F} \sum_i S^i_j x^i - y_j \|_2^2 + \sum_i R(x^i)
\end{align}
Here, $S^i_j$ correspond to multiple sets $c^i_j$ of sensitivity maps, and $x^i$
are multiple sets of images which then have to be combined in a post-processing step.

\section{Calibrationless Parallel Imaging}

\begin{figure}[ht]
\centering
\includegraphics[width=0.48\textwidth]{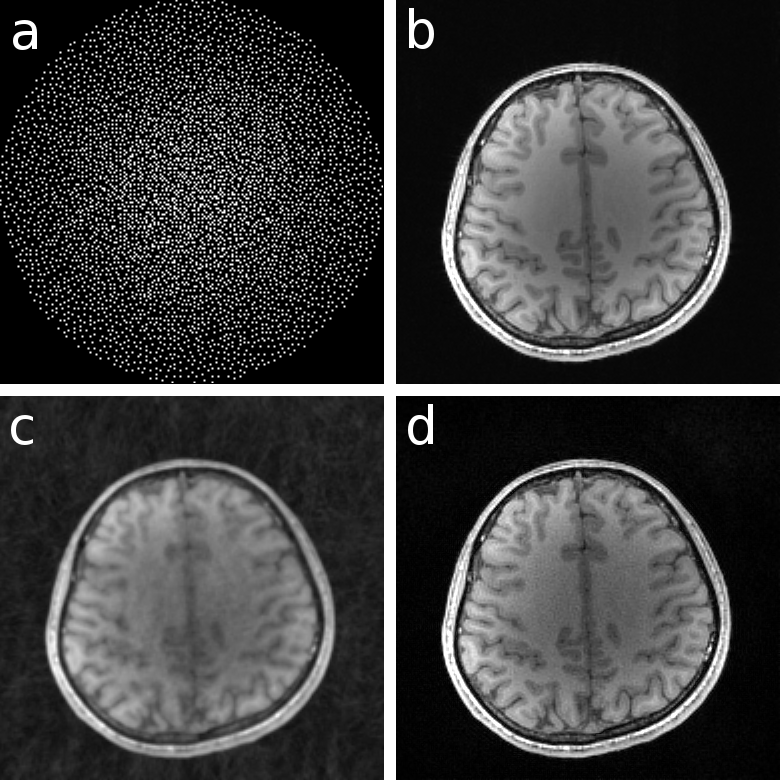}
\caption{Calibrationless parallel imaging using regularized non-linear inversion:
{\bf a:} Variable-density Poisson-disc sampling pattern with six-fold undersampling.
{\bf b:} Reference image reconstructed from fully-sampled data
{\bf c:} Image reconstructed from zero-filled data
{\bf d:} Image reconstructed using regularized non-linear inversion} \label{fig:nlinv}
\end{figure}

For random or non-Cartesian
sampling and when a fully sampled calibration 
region is not available, calibrationless methods
are required.
For example, exploiting the low rankness of the calibration matrix $\mathcal{C}\left\{y\right\}$
missing k-space samples can be recovered
using structured low-rank matrix completion \cite{Shin}:
\begin{align}
	\underset{\hat y}{\operatorname{arg min}} \| y - P \hat y \| \qquad \textrm{with} \quad \operatorname{rank} \mathcal{C}\left\{ \hat y \right\} \leq L
\end{align}
A solution of this problem can be approximated using
Cadzow's algorithm~\cite{Cadzow}, although
this is computationally expensive because it requires
a large SVD in each iteration.
Reconstruction can also be understood directly
as a bilinear inverse problem related to multi-channel blind deconvolution~\cite{JSENSE,Bauer2007,Uecker08}.
The regularized non-linear inversion (NLINV) algorithm~\cite{Uecker08}
jointly estimates the image $x$ and the coil sensitivities $c_j$
based on the formulation as a non-linear inverse problem:
\begin{align}
	\underset{x, c_j}{\operatorname{arg min}} \, & \sum_j \| y_j - P \mathcal{F} \{ c_j x \}\|_2^2 \, + \nonumber\\
		& \qquad \frac{\alpha}{2} ( \sum_j \| W c_j \|_2^2 + \|x\|_2^2 )
\end{align}
Here, $W$ a weighting matrix that penalizes high Fourier coefficients to
exploit the fact that sensitivities are smooth functions. 
Fig.~\ref{fig:nlinv} shows an experimental example.
Inspired by ESPIRiT, this algorithm can be extended 
to multiple images and sets of sensitivities~\cite{Holme17}: 
\begin{align}
	\underset{x, c_j}{\operatorname{arg min}} \, & \sum_j \| y_j - P \mathcal{F} \{ \sum_i c_j^i x^i \}\|_2^2 \, + \nonumber\\
		& \qquad \frac{\alpha}{2} ( \sum_{i,j} \| W c_j^i \|_2^2 + \sum_i \|x^i\|_2^2 )
\end{align}
By introducing new transformed variables $u$ and $v$, and a linear operator $\mathcal{A}$ 
which maps the outer product $X = u v^T$ to $P \mathcal{F} \{ c_j x \}$, it can
be shown that this relaxed version of the algorithm is related to a (convex) structured low-rank matrix recovery 
problem with nuclear-norm regularization~\cite{Recht_SIAMReview_2010}.

\section{Conclusion}

While parallel MRI with known sensitivities is a linear
inverse problem for which results from sampling theory and 
efficient algorithms are readily available,
autocalibrating parallel imaging is related to
multi-variate multi-channel blind deconvolution and
still an active area of research.

\bibliography{blind}

\end{document}